\documentclass{aastex}          
\usepackage{spr-astr-addons}    
\usepackage{multirow,epstopdf}             
\usepackage[Symbol]{upgreek}

\begin{document}
%
\title{The Radio and $\gamma$-ray Variability Analysis of S5 0716+714}

\shorttitle{The Radio and $\gamma$-ray Variability Analysis of S5 0716+714}
\shortauthors{H. Z. Li et al.}

\author{H. Z. ~Li\altaffilmark{1,2,\ddag}, Y. G. Jiang\altaffilmark{3,\dag}, T. F. Yi\altaffilmark{4,\S}, D. F. Guo\altaffilmark{3}, X. Chen\altaffilmark{3}, H. M. Zhang\altaffilmark{5,6}, Q. G. Gao\altaffilmark{1}, F. W. Lu\altaffilmark{1}, J. Y. Ren\altaffilmark{1}}

\altaffiltext{1}{Physics Department, Yuxi Normal University, Yuxi 653100, P. R. China}
\altaffiltext{2}{Key Laboratory for the Structure and Evolution of Celestial Objects, Chinese Academy of Sciences, Kunming 650011, China}
\altaffiltext{3}{Shandong Provincial Key Laboratory of Optical Astronomy and Solar-Terrestrial Environment, Institute of Space Sciences, Shandong University,Weihai, 264209, China}
\altaffiltext{4}{Physics Department, Yunan Normal University, Kunming, Yunnan, 650092, China}
\altaffiltext{5}{GXU-NAOC Center for Astrophysics and Space Sciences, Department of Physics, Guangxi University, Nanning 530004, China}
\altaffiltext{6}{Guangxi Key Laboratory for the Relativistic Astrophysics, Nanning 530004, China}
\altaffiltext{\dag}{jiangyg@sdu.edu.cn}
\altaffiltext{\ddag}{lhz@yxnu.net}
\altaffiltext{\S}{yitingfeng98@163.com}

\begin{abstract}
 We have studied the variability of \object{S5 0716+714} at radio 15 GHz and $\gamma$-ray band using three different methods. A possible periodicity of
$P_{15 GHz}=266.0\pm11.5$ and $P_{\gamma}=344.0\pm16.4$ days are obtained for radio 15 GHz and $\gamma$-ray light curves, respectively. The variability may be related to the intrinsically emission mechanism. The difference between the variability timescales of radio 15 GHz and $\gamma$-ray may be due to that the emission of radio 15 GHz is produced via the synchrotron process, while the $\gamma$-ray is produced by both the SSC and EC processes.
\end{abstract}


\section{Introduction}           
\label{sect:intro}

Blazars are an subclass of active galactic nuclei (AGNs) with large variations in all wavebands. The variability is an important character of blazars, and it is a very useful tool to explore the radiation mechanism of blazars
\citep{Xie94,Xie01,Gaid96,Celo98}.
The confirmed variability time scales can help us \textbf{to} determine the relevant physical parameters of blazars
\citep{Celo98,Lain99,Xie94,Xie01}. Blazars contains two categories, the FSRQ and BL Lac objects.
The object \object{S5 0716+714} (\object{J0721+7120}) is classified as a BL Lac object \citep{Bier81}
with redshift $z=0.31\pm0.08$ \citep{Nils08}. It is one of the
brightest and most well studied radio source. \citet{Wagn95} found
that it is almost always in the active state. 
The observation of Fermi shows that \object{S5 0716+714} is a strong $\gamma$-ray source 
 \citep{Abdo09}.
Diverse time scales variabilities have been discovered in the light
curves of \object{S5 0716+714}. The intra-day
variability (IDV) of \object{S5 0716+714} have
been detected by many authors
 \citep[e.g.][]{Agar16,Chan11,Hees87,Hu14,Quir89,Quir92,Rani10,Wagn96,Wu12,Fosc06,Poon09,Gupt08,Gupt09,Yuan17,Zhang12}.
The short term variability of the object 
also have been
studied in many literatures. \citet{Quir91} studied the correlation
between radio and optical light curve, and found that the variability time scale
changes from 1 to 7 days. 
A periodicity signal with a time scale of 4 days in optical bands was reported by \citet{Heid96}. Based on the 5.3 years optical monitoring, \citet{Qian02} detected a cycle of 10 days. A periodicity about 14 days in I band was obtained by
\citet{Ma04}. \citet{Yuan17} have monitored and analyzed the optical variability of \object{S5 0716+714}. They found that there are the quasi-periods variability on timescales $24.24\pm1.09$, $24.12\pm0.76$, and $24.82\pm0.73$ days in the V, R and I band light curves, respectively. \citet{Rani13a} studied the GeV variability of \object{S5 0716+714}, and found that there are two variable time scales of $\sim$ $75\pm5$ and
$\sim$ $140\pm5$ days.
A variability on timescales from 60 to 90 days 
was reported by \citet{Liao14}. \citet{Wiit09} found that
there exists a very strong signal of a quasi-periodicity of 330 days in
the light curves of \object{S5 0716+714}. Moreover, a 340 days quasi-periodic variability in the $\gamma$-ray light curves was obtained by \citet{Prok17}.
In addition, a 
cycle of 3.3 yr in optical light curves and 5.5-6 yr
in radio light curves were reported by \citet{Rait03}.
Obviously, \object{S5 0716+714} is an extremely active and highly
variable object on diverse time scales.
\citet{Gupt09} studied the IDV of \object{S5 0716+714}
and found that more than one emission mechanism is at work in it. In sec 2, we collect the radio and $\gamma$-ray data. In sec 3, we use three different methods to analyze the variability. Discussion and conclusion is made in sec 4 and sec 5, respectively.

\begin{table*} 
     \caption{The previous detected variability of S5 0716+714.  }
   \label{Tab:publ-works}
  \begin{center}  \tabcolsep0.06in \scriptsize
  \begin{tabular}{|c|c|c|c|c|}  \hline
    Band & Timescale & Duration of the data train &Variability or Periodicity& Reference \\\hline
   Radio and optical	&	1-7 days	&	4 weeks	&V	&Q91	\\\hline
   Optical	&	4 days	&	   	&P&H96	\\\hline
   Optical	&	10 days	&	5.3 years	&P&Q02	\\\hline
   Optical	&	14 days	&	6 years	&P&Ma04	\\\hline
    Optical	&	24.12-24.82 days	&	14 years	&P&Y17	\\\hline
   $\gamma$-ray	&	75 and 140 days	&	4 years	&V&R13	\\\hline
    X-ray	&	330 days	&	12 years	&P&W09	\\\hline
    $\gamma$-ray	&	340 days	&	9 years	&P&P17	\\\hline
    Raido, near-infrared, optical, X-ray, $\gamma$-ray	&	60-90 days	&	1400 days	&V&L14	\\\hline
    Optical	&	3.3year	&	7 years	&V&R03	\\ \hline 
    Radio	&	5.5-6year	&	21years	&P&R03\\\hline
   \end{tabular}\end{center}
   V:Variability; P:Periodicity.\\
   References to Table 1: Q91: Quirrenbach et al., 1991; H96: Heidt \& Wagner 1996; Q02: Qian et al. 2002; M04: Ma et al. 2004; Y17: Yuan et al. 2017; R13: Rani et al. 2013; P17:Prokhorov \& Moraghan 2017; W09: Wiita et al. 2009; L14: Liao et al. 2014; R03: Raiteri et al. 2003.
   \end{table*}

\section{Observation Data and Variability Analysis of the Light Curves}
\label{sect:Obs}

We compiled the variability data of \object{S5 0716+714} at
radio 15 GHz and $\gamma$-ray (0.1-200 GeV).
The 15 GHz data of \object{S5 0716+714} were compiled from the 40 m Telescope at the Owens Valley Radio Observatory (OVRO).
The OVRO starts monitor blazars in 2008 to support the Fermi Gamma-ray Space Telescope (Fermi). Therefore, the sources observed by OVRO have also been monitored by Fermi, which can help us to understand the radiative property of blazars. The interval of the light curve of 15 GHz is over 8.8  years from 8st January 2008 to 15st November 2016.

We downloaded Fermi-LAT photometric data (Pass 8 data) of \object{S5 0716+714}, which covers the observations from 2008 August 6 (Modified Julian Day, MJD 54684) to 2016 December 12 (MJD 57734), from the data server of the Fermi Science Support Center(FSSC) at web\footnote[1]{http://fermi.gsfc.nasa.gov/ssc}.
To minimize systematics, only photons with energies ranging from 100 MeV to 200 GeV were considered in this analysis. A selection of events with zenith angle of 100$^{\circ}$ was applied to avoid the contamination from the Earth limb $\gamma$-rays. Data analysis was performed with the standard analysis tool gtlike/pyLikelihood, which is part of the \emph{Fermi} Science Tools software package\footnote[2]{http://fermi.gsfc.nasa.gov/ssc/data/analysis/scitools/\\overview.html} (ver. v10r0p572). The Pass 8 reprocessed source-class events and the P8R2-SOURCE-V6 set of instrument response functions (IRFs) were used. A correction for the average reduction of the effective area due to pile-up effects as fewer photon events pass the rejection cuts was also considered. This correction is sufficient for integration times longer than a day.

During the process of analysis, photons were selected in a 10$^{\circ}$ circular region of interest (ROI), centered at the position of S5 0716+714. The isotropic background, including the sum of residual instrumental background and extragalactic diffuse $\gamma$-ray background, was modeled by fitting this component at high Galactic latitude (iso-P8R2-SOURCE-V6-v06.txt provided with the \emph{Fermi} Science Tools\footnote[3]{http://fermi.gsfc.nasa.gov/ssc/data/access/lat/\\BackgroundModels.html}). The Galactic diffuse emission model version of "gll-iem-v06.fits" was used. All point sources in the third \emph{Fermi}/LAT source catalog (3FGL, \citet{Acer15}) located in the ROI and an additional surrounding $10^{\circ}$ wide annulus (called "source region") were modeled in the fits, with the spectral parameters kept free only for the brightest sources in the ROI. Depending on the type (spectral or time domain) of analysis, either two or four point sources responded to the criteria of "brightest sources", corresponding to a detection significance in 3FGL having a TS of at least 21.

The light curves of radio 15 GHz and $\gamma$-ray are shown in the
Figure 1. Figure 1 indicates that \object{S5 0716+714} is a
very active object with the variability index of $V_{15 GhZ}=0.69$,
and $V_{\gamma}=0.97$. The variability index
defined by $V=(F_{max}-F_{min})/(F_{max}+F_{min})$ shows the relative emission variability \citep{Fan02}, where $F_{max}$ and $F_{min}$ is the maximal and minimal flux, respectively.
%
In the following sections, we will use the following methods to
analyze the data of \object{S5 0716+714}: the Jurkevich method, the Lomb-Scargle periodogram,
and the red-noise spectra.

\begin{figure}
    \epsscale{1.0} \plotone{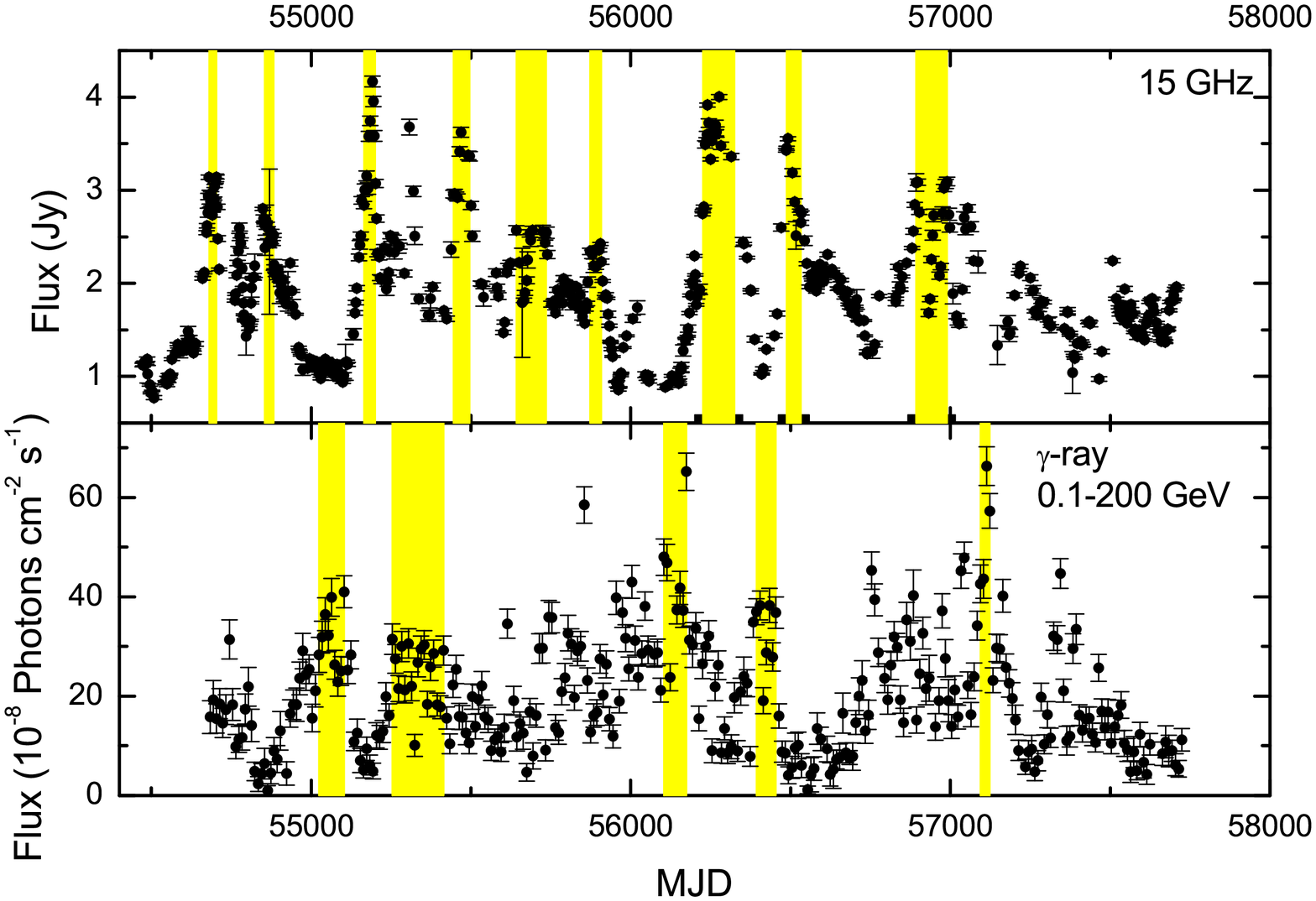}
    \caption{The light curves of S5 0716+714 at 15 GHz and $\gamma$-ray band. The yellow bars in the Figure represent the period during the flux increases.\label{fig1}}
\end{figure}

\section{Analysis Method and Periodicity Results}
\label{sect:Periodicity}
\subsection{The Jurkevich method and results}
The Jurkevich method, described in detail
by \citet{Jurk71}, is based on the expected mean square deviation
and the unequally spaced observations in astronomy observation. It
tests a series of trial periods around which the data are folded.
All data are divided into $m$ groups according to their phases
around each group (bin).
If a trial period equals to a true one, the sums $V^{ 2}_{ m}$ of
all groups would reach its minimum. The sums $V^{ 2}_{ m}$ is calculated following the formula

\begin{equation}
 V^{2}_{m}=\sum_{l=1}^{m} V^{2}_{l},
\label{eq:LebsequeI}
\end{equation}
where $V^{2}_{l}$ is given by the equation
\begin{equation}
  V^{2}_{l}=\sum_{i=1}^{m_{l}} x^{2}_{i}-m_{l}\overline{x}^{2}_{l},
\label{eq:LebsequeII}
\end{equation}
where $x_{i}$ and $m_{l}$ is the individual observation and the number
of observations in the $lth$ group, respectively. In addition, $\overline{x}_{l}$ is defined by the equation
\begin{equation}
  \overline{x}_{l}=\frac{1}{m_{l}}\sum_{i=1}^{m_{l}} x_{i}.
 \label{eq:LebsequeIII}
\end{equation}

For Jurkevich method, \citet{Kidg92} introduced a $f$-test and the
parameter $f$ can be calculated:
\begin{equation}
  f=\frac{1-V_{m}^{2}}{V_{m}^{2}},
\label{eq:LebsequeIf}
\end{equation}
where $V_{m}^{2}$ is the normalized value. The variability behavior
of the light curve can be found from the $V_{m}^{2}$ plot. In
general, $f=0$ implies that there is no periodicity at all; $f \geq
0.5$ implies that the periodicity in the sample is a significant
one; $f \leq0.25$ indicates that there is a weak periodicity. For
$f\leq0.25$, the relationship between the depth of minimum and the
noise in the "flat" section of $V_{m}^{2}$ curve close to the
adopted period needs further test to confirm the significance of the
result. If the absolute value of the relative change of the minimum
to the "flat" section is ten times larger than the standard error of
this "flat" section, the periodicity in the data can also be
considered as significant.

The result of the Jurkevich method is shown in Figure 2. The top and bottom panel of Figure
2 are plotted for the radio 15 GHz and $\gamma$-ray results, respectively. The top panel displays that there are several minimum values of
$V_{m}^{2}$ at the timescale $266.0\pm11.5$, $537.0\pm34.6$, $777.0\pm52.3$ and $1073.0\pm43.9$ days for the radio 15 GHz datasets. This suggests that there are four possible periodicity in the radio 15 GHz light curves.
The parameter $f$ of periodicity $266.0\pm11.5$ days is 0.48, showing that the periodicity of $266.0\pm11.5$ days is a prominent period. The parameter $f$ corresponding to the timescale of $537.0\pm34.6$, $777.0\pm52.3$ and $1073.0\pm43.9$ days all are larger than 0.5, which indicates that they all are significant. Based on a simple Monte Carlo method, a quantitative criterion, the False Alarm Probability (FAP) levels, was calculated with N=10000 \citep{Fan10,Horn86}.  The red line in the top panel of Figure 2 is the FAP levels of 0.05, suggesting that the confidence levels of all the timescale are larger than $95\%$.
Moreover, one can find that the periodicity of $266.0\pm11.5$ days is one half, third and quarter of the timescales of $537.0\pm34.6$, $777.0\pm52.3$ and $1073.0\pm43.9$ days, respectively. This suggests that the periodicity of $266.0\pm11.5$ days is a real period in the 15 GHz light curve of \object{S5 0716+714}, and other timescales of $537.0\pm34.6$, $777.0\pm52.3$ and $1073.0\pm43.9$ days are the astronomical multiple frequencies of the timescale of $266.0\pm11.5$ days.

The bottom panel of Figure 2 shows that the variability timescales in the $\gamma$-ray light curve are $344.0\pm16.4$ and $682.0\pm29.4$ days, respectively. Moreover, one can note that the timescales $682.0\pm29.4$ days
is about two times the timescales $344.0\pm16.4$ days. The parameter $f$ corresponding to those timescales is smaller than 0.25. However, the depth of those
results is significant compared with the ¡°flat¡± section. In addition, the bottom panel also displays that the FAP levels of all timescales are smaller than 0.05. This suggests that there is a tentative periodicity in the $\gamma$-ray light curves.

\begin{figure}
    \epsscale{1.0} \plotone{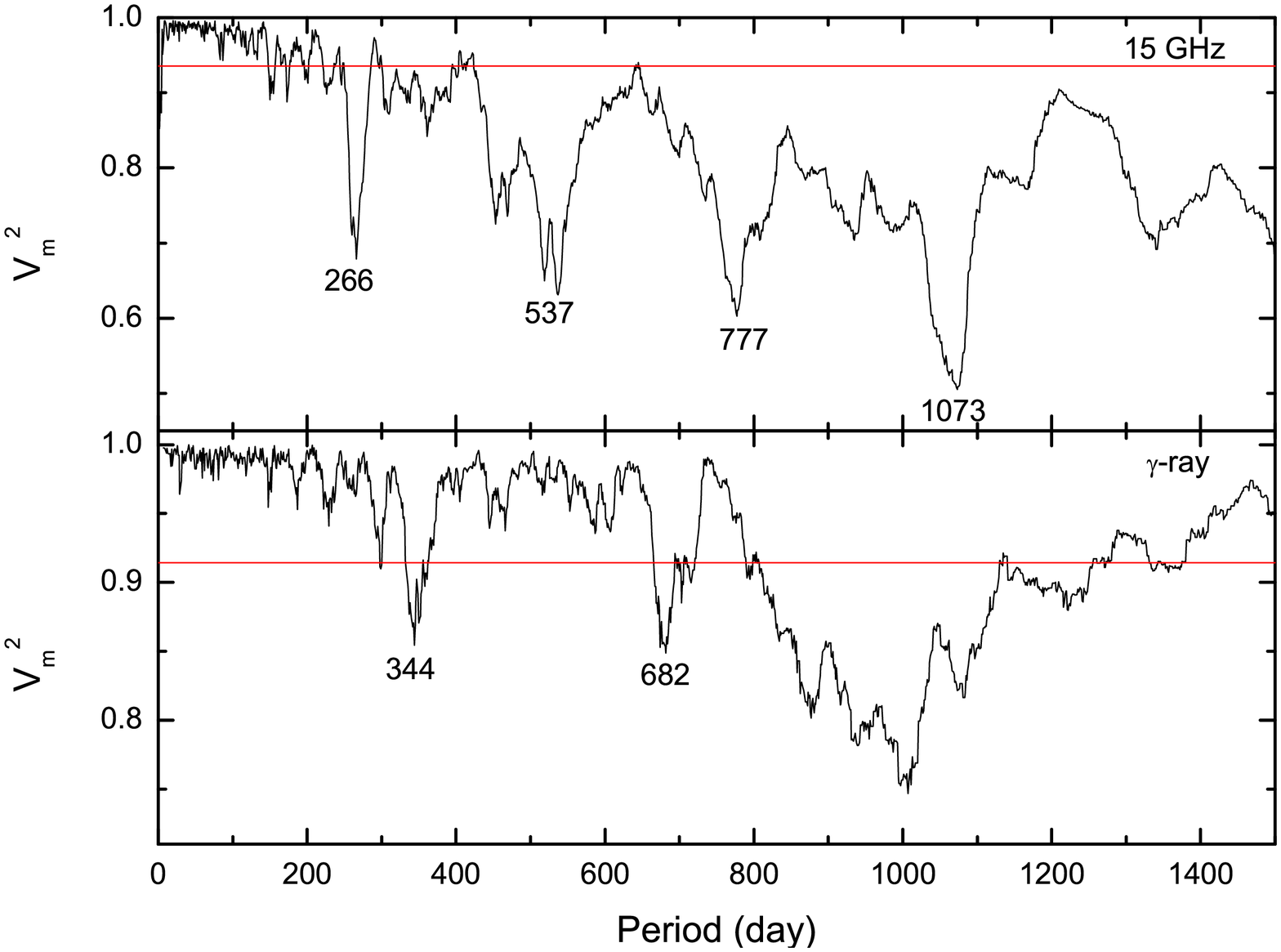}
    \caption{Period analysis results of Jurkevich method for S5 0716+714. The red line is the FAP level of 0.05. \label{fig2}}
\end{figure}

\subsection{The results of the Lomb-Scargle periodogram}
The Lomb-Scargle periodogram is a powerful method to analyze the variability timescale in irregularly sampled
light curves. The algorithm of the Lomb-Scargle periodogram was given by
\citet{Lomb76} and \citet{Scar82} (Lomb-Scargle periodogram). Based
on the Lomb-Scargle periodogram,
the periodogram $P_{X}$ is given by the following formula

\begin{equation}
\begin{aligned}
P_{X}(\omega)= \frac{1}{2}\{\frac{[\Sigma _{i}X(t_{i})cos\omega(t_{i}-\tau)]^{2}}{\Sigma_{i}cos^{2}\omega (t_{i}-\tau)}&\\
+\frac{[\Sigma _{i}X(t_{i})sin\omega(t_{i}-\tau)]^{2}}{\Sigma_{i}sin^{2}\omega
 (t_{i}-\tau)}\},
\end{aligned}
\label{eq:LebsequeIp1}
\end{equation}
where $X(t_{k})$ ($k=0, 1...,N_{0}$) is a times series, $\tau=\frac{1}{2\omega} tan^{-1}[\frac{\Sigma _{i}sin2\omega t_{i}}{\Sigma _{i}cos2\omega t_{i}}]$, and $\omega=2\pi\nu$.
For a power level $z$, the FAP is calculated by \citep{Scar82,Pres94}
\begin{equation}
 p(>z)\approx N\cdot exp(-z),
\label{eq:LebsequeIp}
\end{equation}
where $N$ is the number
of data point \citep{Pres94}.

%

The results derived by the Lomb-Scargle periodogram is presented in the Figure 3. The top and bottom
panel of Figure 3 give the results of 15 GHz and $\gamma$-ray, respectively. The dotted lines in the Figure 3 are the FAP levels of 0.05 which is calculated using the equation (\ref{eq:LebsequeIp}).
From the top panel, three maximum peak at the timescales of $264.0\pm14.0$, $534.5\pm76.1$ and $788.8\pm68.7$ days, suggesting three possible periodicity in the light curve of radio 15 GHz.
Taking account that the timescale of $264.0\pm14.0$ days is about one half and third of the timescale of $534.5\pm76.1$ and $788.8\pm68.7$ days, we consider the timescale of $264.0\pm14.0$ days to be the true variability periodicity in the light curve of radio 15 GHz. This is accord with the results derived by Jurkevich method. The bottom panel of Figure 3 gives the result of $\gamma$-ray derived by the Lomb-Scargle periodogram. This panel shows two possible periods with timescales $345.4\pm16.5$ and $950.0\pm155.8$ days. The bottom panel of Figure 3 also indicates that the FAP levels corresponding to the two periods both are smaller than 0.05. Besides, the timescale $950.0\pm155.8$ days is about three times the timescale $345.4\pm16.5$ days. This is consistent with the results obtained by Jurkevich method and suggests that there is a tentative periodicity in the $\gamma$-ray light curves.

\begin{figure}
    \epsscale{1.0} \plotone{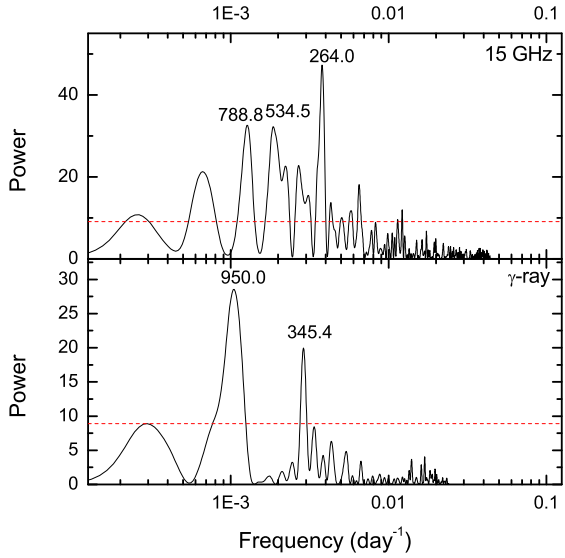}
    \caption{The Lomb-Scargle periodogram results of S5 0716+714. The red line is the FAP level of 0.05. \label{fig3}}
\end{figure}

\subsection{\textbf{The red-noise spectra of the time series}}
Based on the REDFIT38 software \citep{Schu02}, we computed the red noise significance level to compare and further investigate the reliability of foregoing results. Figure 4 plots the red-noise spectra of 15 GHz and $\gamma$-ray wavebands. The top panel of Figure 4 suggests that there is a tentative periodic variability on time scale of $P=269.5$ days ($f=0.00371$) in the light curve of radio 15 GHz, owning significance levels above $99\%$. Moreover, the bottom panel of Figure 4 indicates that the tentative variation periodicity in the light curve of $\gamma$-ray is $P=337.7$ days ($f=0.002961$) with significance levels above $95\%$.
%
The results of 15 GHz and $\gamma$-ray obtained by the REDFIT38 software are consistent with the ones derived by the Lomb-Scargle periodogram and the Jurkevich method.

The signal found in the light curve may be an artificial signal due
to the effect of irregular sampling and seasonal gaps in the
observed light curves. Therefore, we next checked the statistical significance of the $\gamma$-ray result obtained by us using the following
procedure. The first step is to simulate the light curves using the
algorithm described by \citet{Timm95}. Based on a model of a simple
power-law spectrum and the statistical parameter of the
observed data, we generated a series of 10000 simulated
light curves using the algorithm
described by \citet{Timm95} with a power law slope $\beta=2.0$.
The spectral slope $\beta$ of the power spectral density indicates the characteristic of the time series. For the $\beta=1.0$, the light curve is of the flicker type. For $\beta=2.0$, the light curve is of the random walk type which includes red-noise.
Then, the simulated light curves is resampled with the actual sampling function of the observed light curves such that the resampled light curves contain the same irregular sampling and seasonal gaps as the observed data.
Finally, we computed the red noise significance level of the 10000 resampled light curves using the REDFIT38 software. The results suggests that the chance probability of finding the $\gamma$-ray result is smaller than 1 percent. This implies that the tentative variability on time scale about 337.7 days is related to substantial variability in $\gamma$-ray light curves.

\begin{figure*}
  \centering
  \includegraphics  [width=6.0in, angle=0]{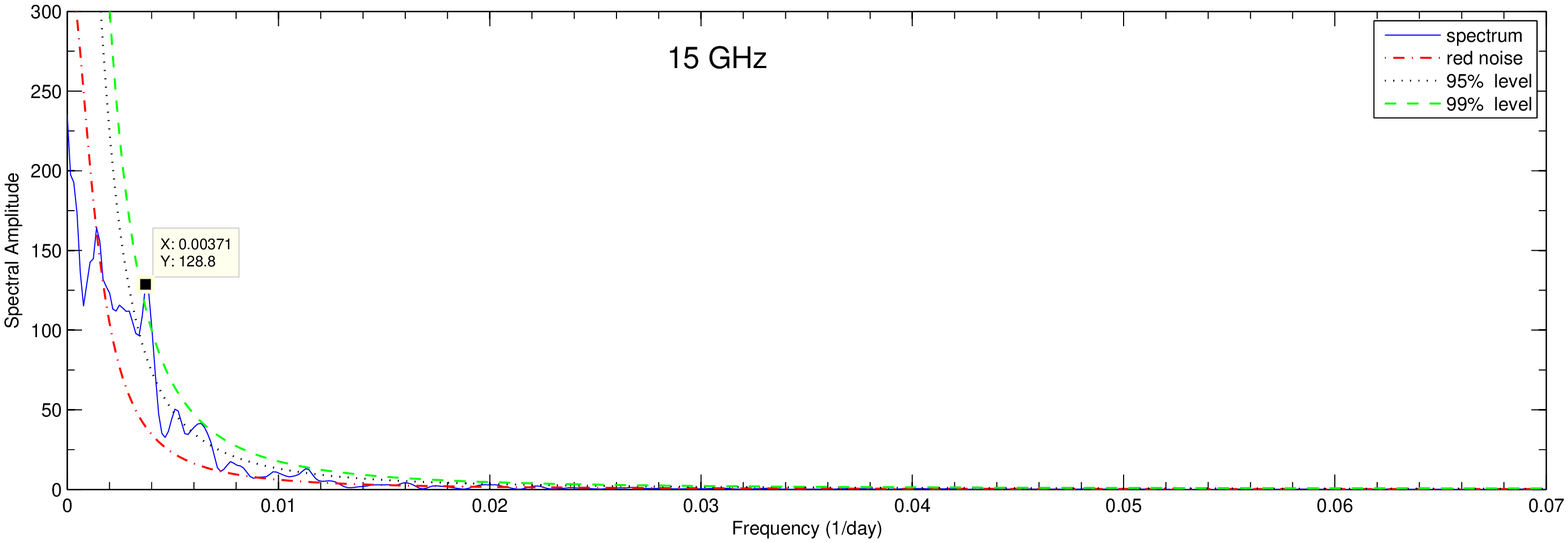}
  \hspace{0.005in}
  \includegraphics  [width=6.0in, angle=0]{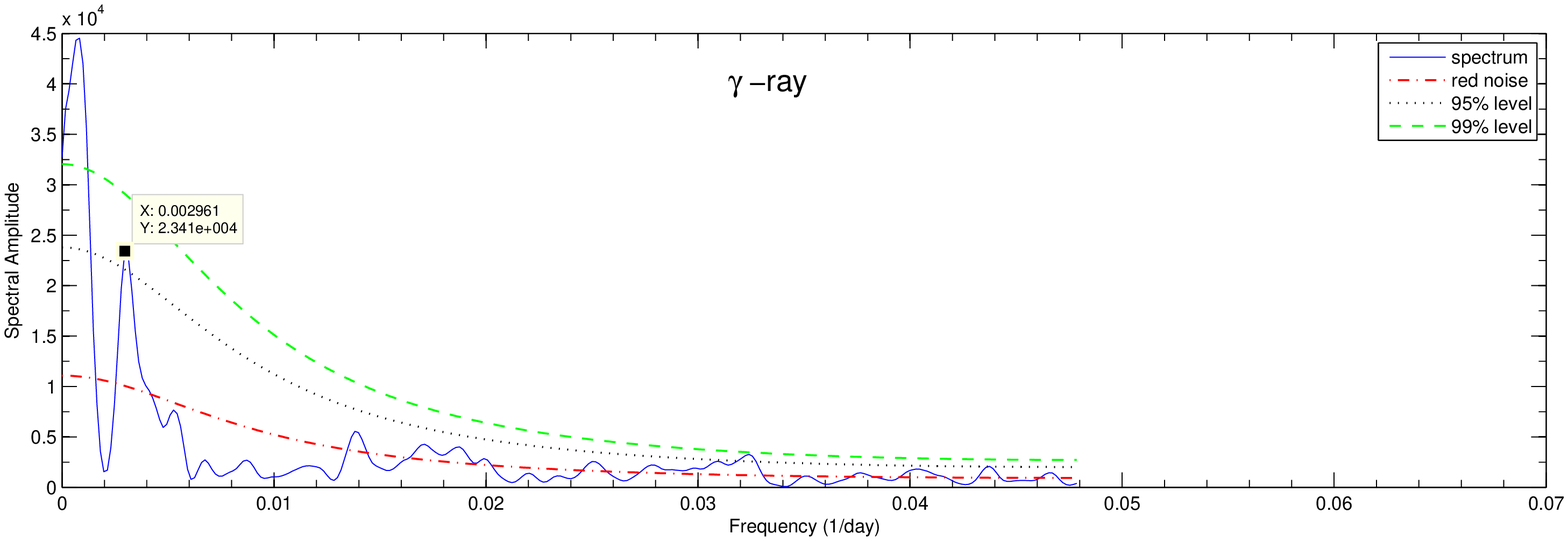}
  \hspace{0.005in}
\caption{The results of S5 0716+714 calculated by the REDFIT38 software. The top and bottom panel show the results of the 15 GHz and $\gamma$-ray band light curves, respectively.}
\end{figure*}


\subsection{Summary}
We have investigated the periodicity of the light curve of \object{S5 0716+714} at radio 15 GHz and $\gamma$-ray wavebands by using three kind of
period analysis techniques. 
Considering the astronomical multiple frequency relationship, a consistent periodicity from 264.0 to 269.5 days was obtained by three methods at radio 15 GHz,
which shows that there is an outburst periodicity on timescale about $P_{15 GHz}=266.0\pm11.5$ days in radio 15 GHz light curve. In addition, a tentative periodicity from 337.7 to 345.4 days is obtained in the $\gamma$-ray band, which implies that there is a tentative outburst periodicity on timescale about $P_{\gamma}=344.0\pm16.4$ days in the $\gamma$-ray band. The cycle of $P_{\gamma}=344.0\pm16.4$ days in the $\gamma$-ray band is accord with the 330 and 340 days timescale reported by \citet{Wiit09} and \citet{Prok17}, respectively.

%

\section{Discussion}
\label{sect:Discussion}
The foregoing analysis suggests that there is a tentative variability periodicity on timescale $P_{15 GHz}=266.0\pm11.5$ and $P_{\gamma}=344.0\pm16.4$ days in the radio 15 GHz, and $\gamma$-ray datasets of \object{S5 0716+714}, respectively. The variability of \object{S5 0716+714} was studied in many literatures \citep{Quir91,Heid96,Qian02,Ma04,Rani13a,Wiit09,Liao14,Rait03}. They found that there are various timescales from 1 days to several years in the light curves of the object. The tentative period on timescale $P_{\gamma}=344.0\pm16.4$ days obtained by us is consistent with the quasi-period of 330 days obtained by \citet{Wiit09}. In addition, \citet{Prok17} suggest that there exists a 340 days quasi-periodic variability in the $\gamma$-ray light curves of \object{S5 0716+714}. This is well consistent with our results. 
On the other hand, \citet{Sand17} have analysed the optical and $\gamma$-ray light curves, and suggests that there is no significant periodicity. However, Figure 3 of \citet{Sand17} reveals that there is an tentative variability on timescales about 300 days in the $\gamma$-ray light curve. This is also agreement with the timescale $P_{\gamma}=344.0\pm16.4$ days obtained by us.
However, there are no obvious signs of 1-7, 4, 10, 14, 24 days, 3.3 and 5.5-6 years period in our results \citep{Quir91,Heid96,Qian02,Ma04,Rait03,Yuan17}. For the timescale of 3.3 and 5.5-6 years obtained by \citet{Rait03}, the duration of the data train used in the paper is too short to confirm them \citep{Kidg92}. 



The radiation of blazars is believed to be the non-thermal radiation from the relativistic jet.
The emission variability of blazars could be due to helical jets or helical structures in jets \citep{Came92,Rieg04,Li09,Li15,Li16}. The emitting flow propagating along a
helical path could cause quasi-periodic variation of Doppler boosting effects, and further leads to emission variability of blazars. Moreover, variability could be result from hydrodynamical instabilities in magnetized jets \citep{Hard99} or from variations in the jet engine which include accretion
disk instabilities \citep{Godf12}, the interaction of the jet plasma with the
surrounding medium, relativistic shocks, etc. For \object{S5 0716+714}, \citet{Liao14} suggest that the scattering of external seed photons, and the synchrotron self-Compton (SSC) process are probably both needed to explain the
$\gamma$-ray emission. The difference between the variability timescales of radio and $\gamma$-ray may be related to the emission processes of them. The radio emission is produced via the synchrotron process, and the $\gamma$-ray emission is produced by both the SSC and external Compton (EC) processes. Thus, the variability timescale of radio 15 GHz and $\gamma$-ray are modulated by the synchrotron emission and the external photon, respectively.

In leptonic scenarios, the lower-energy emission is generated via the synchrotron process, and the high energy emission
can be generated via the inverse Compton (IC)
scattering \citep{Samb96,Tagl03}.
The photons for IC scattering may be from synchrotron radiation (SSC), or the external emission regions of jet (EC) \citep[e.g.][]{Derm93,Derm09,Blaz00}. The short term variability could be related to intrinsic emission processes.
 For the SSC, the variability between the lower-energy emission and the high energy emission should have a good correlation \citep{Ghis97}. For EC, there should be no correlation between them\citep{Ghis97}.
Moreover, the long term variability is usually associated with geometric effects, such as Doppler-modulated due to helical structures and the precession Jet. In the framework of geometric effects and SSC model, the variability should be dominated by the geometric effects, and the flux in both bands should vary together. For EC, the variability timescale of $\gamma$-ray is also modulated by the timescale of external photon which usually is different with the ones of Doppler modulated. Therefore, if the $\gamma$-ray emission is modulated by geometric effects and the timescale of external photon simultaneously, the intrinsic emission variability of radio and $\gamma$-ray also may not necessarily be correlated.
Based on the discrete correlation function (DCF) method \citep{Edel88}, we have analyzed the correlation between the variability of radio 15 GHz and $\gamma$-ray, and the result suggests that there is a weak correlation between them (see Figure 5).
This indicates that the $\gamma$-ray emission of \object{S5 0716+714} can not be produced by the pure SSC process. This supports the results, reported by \citet{Liao14}, that both the SSC and EC processes are needed to explain the $\gamma$-ray emission of
\object{S5 0716+714}.
\begin{figure}
    \epsscale{1.0} \plotone{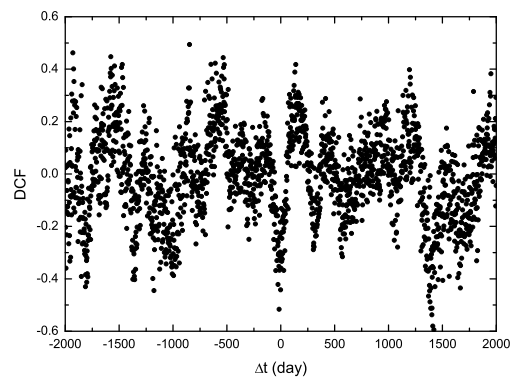}
    \caption{The correlations between the 15 GHz and $\gamma$-ray band of S5 0716+714. \label{fig5}}
\end{figure}

\section{Conclusions}
\label{sect:conclusion}
We presented the historical light curves of \object{S5 0716+714} at radio 15 GHz and $\gamma$-ray band, and investigated the variability 
by using three different analysis methods: the Jurkevich method, the Lomb-Scargle periodogram, and the red-noise spectra. The source exhibits a tentative variability with characteristic timescales of {$P_{15 GHz}=266.0\pm11.5$ and $P_{\gamma}=344.0\pm16.4$ days in the data set of radio 15 GHz and $\gamma$-ray bands, respectively. The variability may be related to the intrinsic emission processes. The difference of the variability timescales between radio 15 GHz and $\gamma$-ray is due to the emission of radio 15 GHz produced via the synchrotron process and the $\gamma$-ray produced by both the SSC and EC processes. The variability timescales of radio 15 GHz and $\gamma$-ray are modulated by the synchrotron emission and the timescale of external photon, respectively.

\acknowledgments
We are grateful to the anonymous referee for useful comments. This work made use of data supplied by Fermi Gamma-Ray Space Telescope project, and the Owens Valley Radio Observatory 40 m monitoring programme, which is
supported in part by NASA grants NNX08AW31G and
NNX11A043G, and NSF grants AST-0808050 and AST-
1109911 (Richards et al. 2011). This work
is supported by the National Natural Science Foundation of China
(11403015,11463001,U1531105), and the Natural Science Foundation of Yunnan Province
(2013FB063), and the Program for Innovative
Research Team (in Science and Technology) in University of Yunnan
Province (IRTSTYN), and the Open Projects of Key Laboratory for the Structure and Evolution of Celestial Objects of Chinese Academy of Sciences (OP201505), and the Young Teachers Program of Yuxi Nurmal University.

%

\end{document}